\documentclass[review,3p]{elsarticle}

\usepackage{epsfig}

\usepackage{amssymb}
\usepackage{amsmath,amssymb,graphicx}
\usepackage{color}
\usepackage{soul}
\usepackage{subfigure}
\usepackage{lineno,hyperref}
\modulolinenumbers[5]
\journal{Physics Letters B}

\begin{document}

\begin{frontmatter}

\title{Analysis of  the quark sector in the 2HDM with  a four-zero Yukawa texture 
 using the most recent data on the CKM matrix}


\author[label1]{O. F\'elix-Beltr\'an}
\ead{olga\_flix@ece.buap.mx}
\author[label1,label2]{F. Gonz\'alez-Canales} 
\ead{felixfcoglz@gmail.com }
\author[label1,label3]{J. Hern\'andez-S\'anchez}
 \ead{jaime.hernandez@correo.buap.mx}
 \author[label4]{S. Moretti}
 \ead{s.moretti@soton.ac.uk}
\author[label3,label5]{R. Noriega-Papaqui}
\ead{rnoriega@uaeh.edu.mx}
\author[label6]{A. Rosado}
\ead{rosado@ifuap.buap.mx}
%%%%%%%%%%%%%%%%%%%%%%%%%%%%%%%%%%%%%%%
\address[label1]{Fac. de Cs. de la Electr\'onica, Benem\'erita Universidad Aut\'onoma de Puebla, Apdo. Postal 542, C. P. 72570 Puebla, Puebla, M\'exico.}
 \address[label2]{Instituto de F\'{\i}sica Corpuscular (CSIC-Universidad de Valencia), Spain.} 
 \address[label3]{Dual C-P Institute of High Energy Physics, M\'exico.}
\address[label4]{School of Physics and Astronomy, University of Southampton, Highfield, Southampton SO17 1BJ, United Kingdom, and Particle Physics Department, Rutherford Appleton Laboratory, Chilton, Didcot, Oxon OX11 0QX, United Kingdom.}
\address[label5]{\'Area Acad\'emica de Matem\'aticas y F\'{\i}sica, 
  Universidad Aut\'onoma del Estado de Hidalgo, 
  Carr. Pachuca-Tulancingo Km. 4.5, C.P. 42184, Pachuca, Hgo. }
 \address[label6]{Instituto de F\'isica, Benem\'erita Universidad Aut\'onoma de Puebla,  Apdo. Postal J-48, C.P. 72570 Puebla, Pue., M\'exico}
%%%%%%%%%%%%%%%%%%%%%%%%%%%%%%%%%%%%%%%%%%%%%%%%%%
\begin{abstract}
 In this letter we analyse, in the context of the general 2-Higgs Doublet Model,  the structure of the 
 Yukawa matrices, $\widetilde{ \bf Y}_{ _{1,2} }^{q}$, by assuming a four-zero texture ansatz for their 
 definition. In this framework, we obtain compact expressions for  $\widetilde{ \bf Y}_{ _{1,2} }^{q}$, which are reduced  to  the Cheng and Sher ansatz with the 
 difference that  they are obtained naturally as a direct consequence of 
 the invariants of the fermion mass matrices. Furthermore, in order to avoid large flavour violating effects coming from charged Higgs exchange, we consider  the main flavour constraints on the off-diagonal terms of Yukawa texture {{$\left( \widetilde{\chi}_{j}^q \right)_{kl}$}} ($k\neq l$). We perform a $\chi^2$-fit based on current experimental data on the quark masses and 
 the Cabibbo-Kobayashi-Maskawa mixing matrix ${ \bf V}_{\rm CKM }$. Hence, we obtain the allowed ranges for 
 the parameters $\widetilde{ \bf Y}_{ _{1,2} }^{q}$ at 1$\sigma$ for several values of $\tan \beta$. The 
 results are in complete agreement with the bounds obtained taking into account constraints on Flavour 
 Changing Neutral Currents reported in the literature.
\end{abstract}

\end{frontmatter}

 %\linenumbers

\section{Introduction} \label{section1}
Now that a Higgs particle has been discovered  at the Large Hadron Collider 
 (LHC)~\cite{Aad:2012tfa,Chatrchyan:2012ufa,Chatrchyan:2013lba}, with properties in very good accordance 
 with the minimal version of Standard Model~(SM)~\cite{Glashow:1961tr,Weinberg:1967tq,Salam:1968rm}, it becomes 
 important to look for extensions of the Higgs sector beyond the SM structure that contain a neutral Higgs 
 boson similar to the one found at the CERN machine. One of the most restrictive experimental results on  
 extensions of the SM is that  Flavour Changing Neutral Currents (FCNCs) must be controlled. The highly 
 experimental suppression for FCNCs should be a test for models with more  than one Higgs multiplet. In 
 particular, in the 2-Higgs Doublet Model~(2HDM)~\cite{Barger:1989fj,Gunion:1989we,Aoki:2009ha}, FCNCs 
 could be avoided through a discrete symmetry $Z_2$. It is well known that there are several versions of 
 this model, known as Type I, II, X and Y (2HDM-I~\cite{Haber:1978jt,Hall:1981bc}, 
 2HDM-II~\cite{Donoghue:1978cj}, 2HDM-X and 
 2HDM-Y~\cite{Barnett:1983mm,Barnett:1984zy,Grossman:1994jb, Akeroyd:1996he,Akeroyd:1998ui,Akeroyd:1994ga}) 
 or \emph{inert}~\cite{Ma:2008uza,Ma:2006km,Barbieri:2006dq,LopezHonorez:2006gr}. The most general version 
 of the 2HDM contains non-diagonal fermionic couplings in the scalar sector implying the generation of 
 unwanted FCNCs. Different ways to suppress FCNCs have been developed, giving rise to a variety of specific 
 implementations of the 2HDM~\cite{Glashow:1976nt,Pich:2009sp,Zhou:2003kd,Kanemura:2005hr,Kanemura:2004cn}. 
 In particular, as it is done in Ref.~\cite{HernandezSanchez:2011ti}, it is possible to analyse the Yukawa 
 matrices through the quark sector phenomenology. The Cheng and Sher {ansatz}~\cite{Cheng:1987rs} has been 
 successfully used to describe the Yukawa  couplings. Several Yukawa textures proposed in 
 literature~\cite{Fritzsch:1977za,Fritzsch:2002ga} have yielded the right description of the Yukawa 
 couplings depending on fermion masses. 
 
 In this paper, we are interested in the Yukawa sector in the context of the general version of the 2HDM  
 considering a four-zero texture fermionic mass matrix~\cite{Fritzsch:1977za,Fritzsch:2002ga}. This Yukawa 
 texture has been studied in 
 Refs.~\cite{DiazCruz:2004pj,DiazCruz:2009ek,BarradasGuevara:2010xs,GomezBock:2005hc,
 HernandezSanchez:2011fq}, which obtained interesting phenomenological results in both charged and neutral 
 Higgs sectors. In this framework, we propose an alternative way to determine the allowed range in the parameter space for the Yukawa matrix elements, taking into account the main flavour physics constraints.  
Besides, considering the current experimental data on the quarks masses and 
  the elements of the Cabibbo-Kobayashi-Maskawa (CKM) matrix, we compute the Yukawa 
 matrices through a
 $\chi^2$-fit over the { $\left( \widetilde{ \bf \chi}_{j}^q \right)_{kl}$} parameters. 
We extract 
 the allowed ranges for  {{$\left( \widetilde{\chi}_{j}^q \right)_{kl}$} as a function of $\tan \beta$ for each 
 like-2HDM (i.e., each of the aforementioned types). The Cheng and Sher structure of the Yukawa couplings 
 is obtained as a natural feature of our fermion mass matrices invariant. Finally, the allowed ranges for 
 the parameters  {{$\left( \widetilde{\chi}_{j}^q \right)_{kl}$} are in accordance with those given in 
 Refs.~\cite{HernandezSanchez:2012eg,HernandezSanchez:2013xj,Crivellin:2013wna}. 
 
%%%%%%%%%%%%%%%%%%%%%%%%%%%%%%%
\section{The Yukawa sector in the 2HDM-III with a four-zero Yukawa texture}\label{section2}
%%%%%%%%%%%%%%%%%%%%%%%%%%%%%%%
 The Yukawa Lagrangian for the quark fields is given by:
 { \begin{equation}\label{2HDMlagrangian}
  \mathcal{L}_{Y} =  \sum_{j=1}^{2} \left( \mathbf{Y}_{j}^{u} \bar{Q}_{L} \tilde{\Phi}_{j} u_{R} 
  + \textbf{Y}_{j}^{d} \bar{Q}_{L} \Phi_{j} d_{R} \right) + \textrm{h.c.}, 
 \end{equation} 
 }
\noindent
where {$\Phi_{j}= ( \phi^{+}_{j}, \phi^{0}_{j} )^{T}$}, denoting the Higgs doublets as
 {$\tilde{\Phi}_{j} = i \sigma_2 \Phi_{j}^{*}$}, and {$\textbf{Y}^{q}_{j}$} with 
 $q = u,d$ are the $3\times3$ complex Yukawa matrices~\cite{Branco:2011iw}. The Yukawa Lagrangian in
 eq.~(\ref{2HDMlagrangian}) has a great deal of free parameters associated with the Yukawa 
 interactions and five Higgs bosons, two of them charged ($H^{\pm}$), two neutral CP-even (scalar) ones 
 ($h^0$ and $H^0$, in increasing order of mass) and one neutral CP-odd (pseudoscalar) state ($A^0$). The 
 mechanism through which the FCNCs are controlled defines the version of the model and the specific 
 emerging phenomenology that can be contrasted with experiments. Before we discuss how FCNCs are 
 controlled in the 2HDM though, we analyse the mass matrix.
 In flavour space, the mass matrix, in general, can be written as
 { \begin{equation}\label{masa-fermiones}
  {\bf M}_{q} =  \frac{1}{ \sqrt{2} } \left( v_{1} \, {\bf Y}_{1}^{q} + v_{2} \, {\bf Y}_{2}^{q} 
  \right), \qquad q = u, d,
 \end{equation} 
 }
 where $v_{1,2}$ are the Vacuum Expectation Values { (VEVs)} of the two Higgs doublet fields. There 
 is no physical restriction on the structure of the mass matrix beyond the fact that the quark masses of 
 different families differ by several orders of magnitude. Consequently, there is no restriction on either 
 Yukawa matrix. The mass matrices {${\bf M}_{q}$} are diagonalised by a biunitary 
 transformation~\cite{Haruo}, 
 { \begin{equation}\label{M:THDM-2}
   {\bf U}_{qL} {\bf M}_{q} {\bf U}_{qR}^{\dagger} =
  \frac{ 1 }{ \sqrt{ 2 } } \left( v_{1}\, \widetilde{ \bf Y}_{ 1 }^{q} 
  + v_2 \, \widetilde{ \bf Y}_{ 2 }^{q} \right) =  {\bf \Delta}_{q} , 
 \end{equation} 
 }
 where {${\bf \Delta}_{q} = \textrm{diag} \left\{ m_{q1}, m_{q2}, m_{q3} \right \}$} and {$\widetilde{ \bf Y}_{j}^{q} = {\bf U}_{qL} {\bf Y}_{j}^{q} {\bf U}_{qR}^{\dagger}$} ($j=1,2$).  This 
 transformation connects the flavour space and the mass space. 
Taking $v^{2} = v_{1}^{2} + v_{2}^{2} = \left( 246.22~\textrm{GeV} \right)^{2}$
 and {$\tan \beta = \frac{ v_{2} }{ v_{1} }$}~\cite{Gunion:1989we,PhysRevD.86.010001}, we have
 { \begin{equation}\label{M:THDM-3}
  \begin{array}{l}
   \left( { \bf \Delta }_{q} \right)_{kl} 
   = \frac{ v \cos \beta }{ \sqrt{ 2 } } \left[ \; 
   \left(\widetilde{ {\bf Y} }_{1}^{q} \right)_{kl}
   + \tan \beta \, \left( \widetilde{ {\bf Y} }_{2}^{q}\right)_{kl}  \; \right],  
  \end{array}
 \end{equation} 
 }
 where $k,l =1,2,3$.  One can see that 
 the off-diagonal elements of the Yukawa matrices in mass space, {$\widetilde{ \bf Y}_{j}^{q}$}, 
 obey the following relation:
 { \begin{equation}\label{off_terms}
  \left( \widetilde{ {\bf Y} }_{1}^{q} \right)_{kl} = 
  - \tan \beta \left( \widetilde{ {\bf Y} }_{2}^{q} \right)_{kl}, \qquad k \neq l.
 \end{equation} 
 }
 Furthermore,  if we require {$\tan \beta$} to be real and 
 positive definite~\cite{Gunion:1989we,Branco:2011iw}, the arguments 
{$ \left( \varphi^{q}_{j} \right)_{kl} = 
 \arg \left( \widetilde{ {\bf Y} }_{j}^{q} \right)_{kl}  $} with $j=1,2$ must satisfy:
 { \begin{equation}\label{CondArg}
  \begin{array}{l}
  \left( \varphi^{q}_{1} \right)_{kl} = \left( 2n + 1 \right) \pi +  
   \left( \varphi^{q}_{2} \right)_{kl}, \qquad k \neq l.
  \end{array}
 \end{equation} 
 }
 The condition in eq.~(\ref{off_terms})  means that the off-diagonal elements of the Yukawa matrices in 
 mass space are parallel or anti-parallel  in the 
 complex plane. In other words, the phases of the off-diagonal elements of the Yukawa 
 matrices  are (anti)aligned.
 In the SM eq.~(\ref{off_terms}) is trivially satisfied, because there exists an alignment between 
 the mass matrix and the corresponding Yukawa matrix. In all 2HDM realisations (Type I, II, X and Y), 
 wherein a discrete symmetry $Z_2$ is imposed, one of the Yukawas is zero. This implies an alignment 
 between the mass matrix and the corresponding Yukawa one. In the Aligned 2HDM (A-2HDM), both Yukawas are 
 aligned in flavour space, which in turn implies an alignment among the mass and Yukawa matrices. In the 
 Minimal Flavour Violating 2HDM (MFV-2HDM), where a (non-discrete) flavour symmetry is imposed, an 
 alignment between the mass matrix and the corresponding Yukawa one is obtained too. In the 2HDM Type III 
 (2HDM-III), with a particular texture form,  eq.~(\ref{off_terms}) is satisfied by construction.
 In this paper we assume a \textit{hierarchical ansatz}, which considers  that the mass matrix and both Yukawa matrices 
 {${\bf Y}^{q}_{j}$} possess the same structure. In particular, we use an Hermitian four-zero texture form and the complete mass matrix inherent to this structure is:
 { 
 \begin{equation}\label{MMM}
  \begin{array}{l}
   {\bf M}_{q} = \left( \begin{array}{ccc}
    0 & C_{q} & 0 \\
    C_{q}^{*} & \tilde{B}_{q} & B_{q} \\
    0 & B_{q}^{*} & A_{q}
   \end{array}\right) = \frac{ v \cos \beta }{\sqrt{2}}  \times \\ 
   \left[  \left( \begin{array}{ccc}
    0 & C_{1}^{q} & 0 \\
    C_{1}^{q\,*} & \tilde{B}_{1}^{q} & B_{1}^{q} \\
    0 & B_{1}^{q\,*} & A_{1}^{q}
   \end{array}\right) 
   + \tan \beta \left( \begin{array}{ccc}
    0 & C_{2}^{q} & 0 \\
    C_{2}^{q\,*} & \tilde{B}_{2}^{q} & B_{2}^{q} \\
    0 & B_{2}^{q\,*} & A_{2}^{q}
   \end{array}\right)  \right].
  \end{array} 
 \end{equation} 
 }
 In the polar form, the off-diagonal elements of the matrices in eq.~(\ref{MMM}) are:
 {$C_{q} = |C_{q}| e^{ i \phi_{c}^{q} }$}, {$B_{q} = |B_{q}| e^{ i \phi_{b}^{q} }$}, 
 {$C_{j}^{q} = |C_{j}^{q}| e^{ i \phi_{cj}^{q} }$}, and 
 {$B_{j}^{q} = |B_{j}^{q}| e^{ i \phi_{bj}^{q} }$}. The Hermitian mass matrices 
 {${\bf M}_{q}$} can be written in terms of a real symmetric matrix {$\bar{\bf M}_{q}$} and a 
 diagonal matrix of phases { ${\bf P}_{q} = \textrm{diag} \left[ 1, e^{i \phi_{c}^{q} }, 
 e^{i \left( \phi_{b}^{q}  + \phi_{c}^{q} \right) }  \right]$} as follows: 
\begin{eqnarray}
 {{\bf M}_{q} = {\bf P}_{q}^{\dagger } \bar{\bf M}_{q} {\bf P}_{q}}.
\end{eqnarray} 
 The eigenvalues $m_{qk}$, $k=1,2,3\,$, of the matrix {$\bar{\bf M}_{q}$} are associated with the 
 quark masses~\cite{Barranco:2010we,Canales:2013cga}. 
 Now, the real symmetric matrix can be brought to a diagonal form by means of the following orthogonal transformation:
\begin{eqnarray}
{\widehat{\bf M}_{q} = {\bf O}_{q} 
 \textrm{diag} \left[ \widehat{m}_{q1}, - \widehat{m}_{q2}, 1 \right] {\bf O}_{q}^{\top}},
\label{Mqtilde}
\end{eqnarray}
 where $\widehat{m}_{q1,q2} = m_{q1,q2}/m_{q3}$  are the ratios of the 
 quark masses, {$\widehat{\bf M}_{q} = \bar{\bf M}_{q} /m_{3q}$} are the normalised mass 
 matrices and {${\bf O}_{q}$} are real orthogonal matrices. In the same way, the 
 Yukawa matrices can be written in polar form as follows: 
\begin{eqnarray}
 {{\bf Y}_{j}^{q} = {\bf P}_{j}^{q \, \dagger } \bar{\bf Y}_{j}^{q} {\bf P}_{j}^{q}}, 
\end{eqnarray}
 where { ${\bf P}_{j}^{q} = \textrm{diag} \left[ 1, e^{i \phi_{cj}^{q} }, 
 e^{i \left( \phi_{bj}^{q}  + \phi_{cj}^{q} \right) }  \right]$} and {$\bar{\bf Y}_{j}^{q}$} is a 
 real symmetric matrix. 
Otherwise, the unitary matrices in eq.~(\ref{M:THDM-2}) satisfy the condition 
 {${\bf U}_{qL} ={\bf U}_{qR} = {\bf U}_{q} $} and can be written as 
 {${\bf U}_{q} = {\bf O}_{q}^{\top} {\bf P}_{q}$}~\cite{Barranco:2010we,Canales:2013cga}. 
 Since eq.~(\ref{Mqtilde}), one can obtain the following invariants of the real symmetric mass matrices: 
 {$\textrm{Tr} \left\{ \widehat{\bf M}_{q} \right\}$}, 
 {$\textrm{Tr} \left\{ \widehat{\bf M}_{q}^{2} \right\}$} and 
 {$\textrm{Det} \left\{ \widehat{\bf M}_{q} \right\}$}, respectively, given by
 { 
 \begin{equation}\label{Invariantes}
  \begin{array}{l} \vspace{.3mm}
   c_{q}^{2} \, a_{q} = \widehat{m}_{q1} \, \widehat{m}_{q2}, \quad
   \widetilde{b}_{q} + a_{q} = 1 + \widehat{m}_{q1} - \widehat{m}_{q2}, \\ \vspace{.3mm}
   c_{q}^{2} + b_{q}^{2} - \widetilde{b}_{q} a_{q} = \widehat{m}_{q1} \, \widehat{m}_{q2} 
   - \widehat{m}_{q1} + \widehat{m}_{q2},
  \end{array}
 \end{equation} 
 }
 where $a_{q} = A_{q}/m_{q3}, \, \widetilde{b}_{q} = \widetilde{B}_{q}/m_{q3}, b_{q} = \left| B_{q} \right|/m_{q3} , \, c_{q} = \left| C_{q} \right|/m_{q3}$.  If  we define $a_q = 1-\delta_q$,  the mass matrices {$\widehat{ {\bf M} }_{q}$} can be parameterised as~\cite{Barranco:2010we}:
 { 
 \begin{equation}\label{TwoCeros}
  \widehat{ \bf M}_{q} = 
  \left(\begin{array}{ccc}
   0 & \sqrt{ \frac{ \widehat{m}_{q1} \widehat{m}_{q2} }{ \left(  1 - \delta_{q} \right) } } & 0 \\
   \sqrt{ \frac{\widehat{m}_{q1} \widehat{m}_{q2} }{ \left( 1 - \delta_{q} \right) } } &
   \widehat{m}_{q1} - \widehat{m}_{q2} + \delta_{q} &
   \sqrt{\frac{ \delta_q }{\left( 1 - \delta_{q} \right)} f_{q1} f_{q2} } \\
    0 &\sqrt{\frac{\delta_{q}}{\left( 1 - \delta_{q} \right) } f_{q1} f_{q2}  }  & 1- \delta_{q}
  \end{array}\right),
 \end{equation} 
 }
 where $f_{q1} = \left( 1 - \widehat{m}_{q1} - \delta_{q} \right)$ and 
 $f_{q2} = \left( 1 + \widehat{m}_{q2} - \delta_{q} \right)$. 
One can see from eq.~(\ref{MMM}) that the quark mass matrix  has six free parameters and using 
 eq.~(\ref{Invariantes}) we can fix three of them, which can be  the phases $\phi_{b}^{q}$, $\phi_{c}^{q}$ 
and the parameter $\delta_{q}$. Using the strong hierarchy in the masses of the quark families, $m_{q3} >> m_{q2} > m_{q1}$, we constrain $1-\delta_{q}$ to be very close to unity. Keeping in mind this idea and
 following the analysis of Refs.~\cite{Barranco:2010we,Mondragon:1998gy,Canales:2012dr}, one can obtain the constraint  $0 < \delta_{q} < 1 - \widehat{m}_{1q}$.   
%%%%%%%%%%%%%%%%%%%%%%%%%%%%%%%%%%%%
 In this work the quark mass matrices have been normalised with 
 respect to the heaviest quark mass. We consider   the mass quarks ratios because our results are more stable 
 at the scale  of $Q= m_{H^\pm}$  when the running quark masses are considered~\cite{PhysRevD.86.010001,Xing:2007fb}. 
%%%%%%%%%%%%%%%%%%%%%%%%%%%%%%%%%%%%%%%%
 The orthogonal real matrices {${\bf O}_{q}$} can also be written in terms of the  eigenmasses ratios when 
 {$\bar{\bf M}_{q}$} is calculated, such as in the  Ref.~\cite{Barranco:2010we}:
 { 
 \begin{equation}\label{ortogonal_O}
   {\bf O}_{q} = 
  \left(\begin{array}{ccc}
     \left[ \frac{ \widehat{m}_{q2} f_{q1} }{ {\cal D}_{q1} } \right]^{ \frac{1}{2} } &
   - \left[ \frac{ \widehat{m}_{q1} f_{q2} }{ {\cal D}_{q2} } \right]^{ \frac{1}{2} } &
     \left[ \frac{ \widehat{m}_{q1} \widehat{m}_{q2} \delta_{q} }{ {\cal D}_{q3} }
     \right]^{ \frac{1}{2} } \\
     \left[ \frac{ \widehat{m}_{q1} \left( 1 - \delta_{q} \right) f_{q1} }{ {\cal D}_{q1} }
     \right]^{ \frac{1}{2} } &
     \left[ \frac{ \widehat{m}_{q2} \left( 1 - \delta_{q} \right) f_{q2} }{ {\cal D}_{q2} }
     \right]^{ \frac{1}{2} } &
     \left[ \frac{ \left( 1 - \delta_{q} \right) \delta_{q} }{ {\cal D}_{q3} } 
     \right]^{ \frac{1}{2} } \\
   - \left[ \frac{ \widehat{m}_{q1} f_{ q2 } \delta_{q} }{ {\cal D}_{q1} }
     \right]^{ \frac{1}{2} } &
   - \left[ \frac{ \widehat{m}_{q2} f_{q1} \delta_{q} }{ {\cal D}_{q2} }
     \right]^{ \frac{1}{2} } &
     \left[ \frac{ f_{q1} f_{q2} }{ {\cal D}_{q3} }\right]^{ \frac{1}{2} }
   \end{array} \right),
  \end{equation} 
  }
  with
\begin{eqnarray}
  \mathcal{D}_{q1} &=& \left( 1 - \delta_{q} \right) \left( \widehat{m}_{ q1 } 
  + \widehat{m}_{ q2 } \right) \left( 1 - \widehat{m}_{ q1 } \right) , \\ \nonumber 
  \mathcal{D}_{q2} &=& \left( 1 - \delta_{q} \right) \left( \widehat{m}_{ q1 } 
  + \widehat{m}_{ q2 } \right) \left( 1 + \widehat{m}_{ q2 } \right),   \\ \nonumber
  \mathcal{D}_{q3} &= & \left( 1 - \delta_{q} \right) \left(1 - \widehat{m}_{ q1 } \right) 
  \left( 1 + \widehat{m}_{ q2 } \right) .
\end{eqnarray}
 When the Yukawa matrices are represented by a four-zero texture, these matrices in mass space have the following form:
\begin{eqnarray}
  \widetilde{\bf Y }_{j}^{q} = m_{q3} 
  {\bf O}_{q}^{\top} {\bf Q}_{j}^{q \, \dagger} \, \widehat{\bf Y }_{j}^{q} \,  {\bf Q}_{j}^{q} 
  {\bf O}_{q},
\label{Yuk_Rot-2}
\end{eqnarray} 
 where {$\widehat{\bf Y }_{j}^{q} = \bar{\bf Y }_{j}^{q}/m_{q3}$} is a real symmetric matrix normalised with respect to the heaviest quark, and 
 {${\bf Q}_{j}^{q} = {\bf P}_{j}^{q} {\bf P}_{q}^{\dagger} = \textrm{diag} 
 \left[ 1, e^{i \varphi_{cj}^{q}}, e^{ i \left( \varphi_{bj}^{q} + \varphi_{cj}^{q} \right)} \right]$}
 with {$\varphi_{bj}^{q} = \phi_{bj}^{q} - \phi_{b}^{q} $} and 
 {$\varphi_{cj}^{q} = \phi_{cj}^{q} - \phi_{c}^{q}$} (these phases are the difference between phases coming from Yukawa and mass matrices).
 After some algebra, the matrices { $\widetilde{\bf Y }_{j}^{q}$}, eq.~(\ref{Yuk_Rot-2}), can 
 now be written in the following compact and generic form:
 \begin{eqnarray}\label{ChenSher-2}
  \left( \widetilde{Y}_{j}^{q} \right)_{kl} =
   \dfrac{ \sqrt{ \; m_{qk} \, m_{ql} \; } }{v} \;
   \left( \widetilde{\chi}_{j}^{q} \right)_{kl}, \qquad k,l = 1,2,3.
 \end{eqnarray} 
   These expressions  correspond to the Cheng and Sher ansatz, which are obtained in previous works  with others similar parameterisations~\cite{HernandezSanchez:2011ti,Cheng:1987rs,DiazCruz:2004pj}.  The  coefficients $ \widetilde{\chi}_{j}^{q} $  are functions of the Yukawa matrix parameters {$A_{j}^{q}$}, {$B_{j}^{q}$},  {$C_{j}^{q} $,  
{$\widetilde{B}_{j}^{q}$}} (see  eq.~(\ref{MMM})), and the mass matrix parameter $\delta_q$.
%%%%%%%%%%%%%%%%%%%%%%%%%
%answer reviewer 3
We want to point out that the parameters $ \widetilde{\chi}_{j}^{q} $ have an additional dependence on the charged Higgs boson mass when the couplings of the $H^\pm$ state with fermions are considered in the flavour physics processes and the constraints for those parameters are  obtained (see  Refs.~\cite{HernandezSanchez:2012eg,HernandezSanchez:2013xj,Crivellin:2013wna}). 
%%%%%%%%%%%%%%%%%%%%%%%%%%
In order to determine the numerical value of these coefficients for each 
quark sector, we have to know the values of $\widehat{m}_{qj}$ and $\delta_{q}$ at the scale $Q = m_{H^\pm} $ for 
   90 GeV$ \leq m_{H^\pm} \leq  500$ GeV:
 { 
 \begin{eqnarray}
   \widehat{m}_{u} &\simeq & \left( 1.73 \pm 0.75 \right) \times 10^{-5},  \,\,\,
   \widehat{m}_{c} \simeq \left( 3.46 \pm 0.43 \right) \times 10^{-3}, \\ \nonumber
   \widehat{m}_{d} & \simeq & \left( 1.12 \pm 0.007 \right) \times 10^{-3},  \,\,\,
   \widehat{m}_{s} \simeq \left( 2.32 \pm 0.84 \right) \times 10^{-2}.
 \end{eqnarray} 
 }
 The values of the running quark masses at scale  $Q = m_{H^\pm}$ were calculated with the {\tt RunDec} 
 program~\cite{Chetyrkin:2000yt}. 
Furthermore,  the parameters $\delta_{q}$ can be determined  with the help of 
the quark flavour mixings. 
 Now, in order to  contrast our theoretical expression of  the CKM matrix with the recent 
 experimental data, via a $\chi^{2}$ fit,
 we should give  the theoretical expressions for the elements of  the quark mixing matrix, which is obtained by the definition of  ${\bf V}_{_{\rm CKM}}$: 
 { 
 \begin{equation}\label{MatCKM}
  {\bf V}_{_{\rm CKM}}^{^{th}} = {\bf U}_{u} {\bf U}_{d}^{\dagger} =
  {\bf O}_{u}^{ \top } {\bf P}^{(u-d)} {\bf O}_{d},
 \end{equation} 
 }
 where {${\bf O}_{u,d}$} are the real orthogonal matrices given in eq.~(\ref{ortogonal_O}) and 
 {${\bf P}^{(u-d)}= \textrm{diag} \left[ 1, e^{i\phi_{1}}, e^{i( \phi_{1} + \phi_{2} )} \right] $}
 with {$\phi_{1} = \phi_{c}^{u} - \phi_{c}^{d}$} and {$\phi_{2} = \phi_{b}^{u} - \phi_{b}^{d}$}
  (these phases are the difference between phases coming from the up- and down- mass matrices).  
 Now,  we make a $\chi^{2}$ fit through the following function~\cite{Canales:2013cga,Canales:2012ix}:
 {
 \begin{equation}
  \chi^{2} = 
  \sum_{m = d,s,b} \frac{ \left( \left| V_{um}^{^{th}} \right| - \left| V_{um} \right| \right)^{2} }{ 
   \sigma_{V_{um}}^{2} } 
  + \frac{ \left(\mathcal{J}^{^{th}}_q - \mathcal{J}_q  \right)^2}{\sigma_{{\mathcal{J}_q}}^2},
 \end{equation}
 } 
 where the terms with super-index $``th"$ are given in eq.~(\ref{MatCKM}) and the quantities without 
 super-index are given by the experimental data with uncertainty 
 $\sigma_{V_{kl}}^{2}$~\cite{PhysRevD.86.010001}:
 { 
 \begin{eqnarray}
   \left| V_{ud} \right| & = & 0.97427 \pm 0.00015, \;
   \left| V_{us} \right| = 0.2253  \pm 0.007, \\ \nonumber
   \left| V_{ub} \right| & = & 0.00351 \pm 0.00015,\;
   \mathcal{J}_{q}= (2.96\pm 0.18)\times 10^{-5}.
 \end{eqnarray} 
 } 
The CKM  matrix, eq.~(\ref{MatCKM}), has four free parameters $\phi_{1}$, $\phi_{2}$, 
 $\delta_{u}$ and $\delta_{d}$. But, in Ref.~\cite{Canales:2013cga} it is shown that, if the quarks mass 
 matrices are represented through a matrix with four-zero texture,  the best values for the 
 $\chi^{2}$ function are  obtained when {$\phi_{2} = 0$} and for a large value of {$\cos 
 \phi_{1}$}. Therefore, without loss of generality, we perform a $\chi^{2}$ fit with the 
 following values for the phases given in eq.~(\ref{MatCKM}): $\phi_{1} = \pi/2$ and  
$\phi_{2} = 0$~\cite{Barranco:2010we,Canales:2013cga,Canales:2012dr,Canales:2012ix}. 
Then, the 
 $\chi^{2}$ function has only two effective free parameters and  the best  values that we obtain for the free parameters 
 $\delta_{u}$ and $\delta_{d}$ at 1$\sigma$ are: 
 { 
 \begin{equation}\label{DeltasQ}
  \delta_{u} = \left( 5.14_{-2.2}^{+4.0}  \right) \times 10^{-2} \quad % \textrm{and} \quad
  \delta_{d} = \left( 3.36_{-1.71}^{+3.32} \right) \times 10^{-2}.
 \end{equation} 
 }
 Furthermore, from the values for the quark mass ratios given above, the parameters of eq.~(\ref{DeltasQ}) and the 
 moduli of the entries of the quark mixing matrix, we have at 1$\sigma$:
% { $\left| \left( V_{ _{CKM} }^{ ^{th} } \right)_{kl} \right|$ } and 
 %the Jarlskog invariant take the follows values, at 1$\sigma$,
 %
 { \begin{equation}
   \left| V_{ _{{\rm CKM }} }^{ ^{th} } \right|_{1 \sigma} =
  \left( \begin{array}{ccc} \vspace{1mm}
   0.97427 \pm 0.00023 & 0.22533_{-0.00096}^{+0.0010}  & 0.00351_{-0.00023}^{+0.0072} \\ \vspace{1mm}
   0.22520 \pm 0.00100 & 0.97324_{-0.00023}^{+0.00058} & 0.0458_{-0.010}^{+0.0033} \\ \vspace{1mm}
   0.00894_{-0.00069}^{+0.0016} & 0.0451_{-0.010}^{+0.048} & 0.998944_{-0.00016}^{+0.00042}
   \end{array}   \right)
 \end{equation} 
 }
 as well as the Jarlskog invariant with the value
\begin{eqnarray}
{ \mathcal{J}^{^{th}}_{ q } = \left( 2.96 \pm 0.28 \right) \times 10^{-5}},
\end{eqnarray} 
 which  is in good agreement with experimental data~\cite{PhysRevD.86.010001}.
 Correspondingly, the numerical values of the normalised symmetric mass matrices 
 {$\widehat{M}_{d}$} and {$\widehat{M}_{u}$} given in eq.~(\ref{TwoCeros}), 
 at 1$\sigma$, are:
 {
 \begin{equation}\label{Md:valores}
  \widehat{M}_{d} = 
  \left( \begin{array}{ccc}\vspace{1mm}
   0 & \left( 5.18_{-0.056}^{+0.30} \right) \times 10^{-3}  & 0 \\ \vspace{1mm}
   \bullet & 
   \left( 1.16_{-1.71}^{+3.32} \right) \times 10^{-2} & 0.183_{-0.054}^{+0.070} \\ \vspace{1mm}
   \bullet &\bullet  & 0.966_{-0.033}^{+0.018}
  \end{array}   \right),
 \end{equation} 
 }
 { 
 \begin{equation}\label{Mu:valores}
  \widehat{M}_{u} =
  \left( \begin{array}{ccc} \vspace{1mm}
   0 & \left( 2.45_{-0.335}^{+0.63} \right) \times 10^{-4}  & 0 \\ \vspace{1mm}
   \bullet  & 
   \left( 4.82_{-2.18}^{+3.96} \right) \times 10^{-2} & 0.221_{-0.050}^{+0.067} \\ \vspace{1mm}
   \bullet & \bullet & 0.949_{-0.040}^{+0.022}
  \end{array}   \right).
 \end{equation}   
 }
With these results  we can establish a hierarchical four-zero texture ansatz for the quark mass matrices ${\bf M}_{q}$, namely
 $|A_q| >> |\tilde{B}_b|, \, |B_b|, \, |C_b|$, which  is not necessary imposed to
 the Yukawa matrices ${\bf Y}^{q}$. However, when  this ansatz and the additional criterion $|A^q_j|, m_{q3}, m_{q2} >> m_{q1}$ for the Yukawa matrices are assumed, the coefficients $\widetilde{\chi}_{j}^{q} $
 given in eq.~(\ref{ChenSher-2}) have the same form as those reported in Ref.~\cite{DiazCruz:2004pj}. Note that,  in our parameterisation, we do not consider any assumptions about the Yukawa matrices. However, since  eq.~(\ref{ChenSher-2}) is the same for both parameterisations, we can get the same structure of Yukawa couplings $H^\pm f_{u_i} f _{d_j}$ and all phenomenological consequences can be applied.
Therefore, 
 we obtain the same  Lagrangian of the charged  Higgs coupled with quarks given by~\cite{HernandezSanchez:2012eg}:
\begin{eqnarray}
{\cal{L}}^{\overline{f}_{i}f_{j}H^{+}}&=&- \{ \frac{\sqrt{2}}{v} \overline{u}_{i} (m_{ d_j} X_{lj}P_{R} + m_{u_i} Y_{ij} P_{L} ) d_{j} H^{+} + {\textrm{H.c.}}\}, 
\end{eqnarray}
with
\begin{eqnarray}
X_{ij}&=&\sum_1^3 (V_{{\rm CKM }})_{il} \bigg[  X \frac{m_{d_l}}{m_{d_j}} \delta_{lj} -  \frac{ f(X) }{ \sqrt{2} } \sqrt{\frac{m_{d_l} }{m_{d_j} } } \chi^{d}_{lj} \bigg] , \\ 
Y_{ij}&=&\sum_1^3 \bigg[  Y  \delta_{il} -  \frac{ f(Y) }{ \sqrt{2} } \sqrt{\frac{m_{u_l} }{m_{u_i} } } \chi^{u}_{il} \bigg] (V_{{\rm CKM }})_{lj} ,
\end{eqnarray}
where $\chi^{f}_{ij}$ was introduced in eq.~(\ref{ChenSher-2}), $f(x)= \sqrt{1-x^2}$, the parameters $X$, $Y$ are real and   can be related to $\tan \beta$ or $\cot \beta$, according to the model considered, namely 2HDM Type I, II, X and Y (see the analysis of Refs.~\cite{HernandezSanchez:2012eg,HernandezSanchez:2013xj}). 

%%%%%%%%%%%%%%%%%%%%%%%%%%%%%%%
\section{Fit of Yukawa matrices and the input parameters free from violating effects }\label{section3}
%%%%%%%%%%%%%%%%%%%%%%%%%%%%%%%

In our model, we cannot assume that only the Standard Model $W$-exchanged charged current contributes to the observables used to determine the CKM  entries, because the fermion-Higgs couplings are not aligned with the fermion mass matrices and  there are  flavour violating contributions coming
from charged Higgs exchange. Then, we should guarantee that our model  is free from flavour violating effects that exceed the current bounds. 
In particular, meson-physics processes  allow to determine several elements of the $V_{{\rm CKM }}$ matrix. Some of those processes are very sensitive to charged Higgs boson exchange, like the leptonic decays $B \to \tau \nu_\tau$, $D \to \mu \nu$, $D_s \to  \ell \nu$, the semileptonic transition $ B \to D \tau \nu_{\tau}$, the inclusive decay $B\to X_s \gamma$, $B_0-B_0$ mixing, $B_s\to \mu^+ \mu^- $ and the radiative decay $Z \to b \bar{b}$, all of which are analysed in~\cite{HernandezSanchez:2012eg}.  
In order to obtain a parameter space consistent with the current experimental results, we use the main flavour constraints and recent analysis of Ref.~\cite{HernandezSanchez:2012eg},
 in particular, for the off-diagonal terms of Yukawa texture given in eq.~(\ref{ChenSher-2}), which leads to flavour violating effects. Taking into account the analysis previously mentioned, we can assume that the diagonal terms of the Yukawa texture take values of  $O(1)$ (this case has been  studied and can avoid the flavour physics constraints), then we can scan the parameters space of the model isolating the surviving off-diagonal terms of the Yukawa texture, obtaining a average range for $(\widetilde{\chi}_{n}^{f}  )_{ij}$ ($i\neq j$):
 {
 \begin{equation}
  \begin{array}{l}
  -0.06 \leq \left( \widetilde{\chi}_{n}^{d}  \right)_{23} \leq 0.3, \;
  -0.3 \leq \left( \widetilde{\chi}_{n}^{u}  \right)_{23}\leq 0.5, \; 
  -0.1 \leq \left( \widetilde{\chi}_{n}^{q}  \right)_{13} \leq 0.1.
  \end{array}
  \label{eq:rangechis}
 \end{equation}
 } 
Moreover,  perturbativity, electroweak  and unitarity constraints are imposed~\cite{Cordero-Cid:2013sxa}. Besides, the different scenarios with a small charged Higgs mass are consistent with the current measurements from flavour and electroweak physics~\cite{HernandezSanchez:2012eg,HernandezSanchez:2013xj,Crivellin:2013wna}.

%%%%%%%%%%%%%%%%%%%%%%%%%%%%%%%%%%%%%%%
% the last section
%%%%%%%%%%%%%%%%%%%%%%%%%%%%%%%%%%%%%%%%

Now we proceed with the fit to the Yukawa texture: by considering the quark mass ratios and the values of  $\delta_u$ and $ \delta_d$ given in eq.~(\ref{DeltasQ}),
 we compute the other values of the coefficients 
 {$\left( \widetilde{ \bf \chi }_{j}^{q} \right)_{kl}$}  for both quark sectors. 
Considering the free parameters 
 { $ \left( \widetilde{\chi}^{q}_{j} \right)_{kl} $} 
 to be real, the phases in eq.~(\ref{ChenSher-2}) satisfy the conditions: 
 $\varphi_{bj}^{q} = \phi_{bj}^{q} - \phi_{b}^{q} = 0$ and 
 $\varphi_{cj}^{q} = \phi_{cj}^{q} - \phi_{c}^{q} = 0$. Namely, we have an alignment 
 between the phases of the mass matrix with the one of the Yukawa matrices.
 Furthermore, the entries of the Yukawa matrices, right side of eq.~(\ref{MMM}), normalised with respect  to the 
 heaviest quark, can be written in terms of the parameters  { $ \left( \widetilde{\chi}^{q}_{j} \right)_{kk} $}  $( k=1,2,3)$.
Hence, in order to find the allowed regions for the free parameters 
 { $ \left( \widetilde{\chi}^{q}_{j} \right)_{kk} $}, taking in account the inputs of eq.~(\ref{eq:rangechis}), we define the 
 { $\chi^{2}_{ _{ M_q } }$} function   
 as:
 %a new  $\chi^{2}$ function as
 { 
 \begin{equation}\label{NewChi}
  \begin{array}{l}
  \chi^{2}_{ _{ M_q } } =  
  \sum_{k}^{3}
   \frac{ \left[ \left(  M_{q}^{fit} \right)_{kk} - \left(  M_{q}^{th} \right)_{kk} \right]^{2} }{
    \left( \sigma_{ \left( M_{q} \right)_{kk} } \right)^{2}  } 
    + \frac{1}{2} \sum_{k \neq l}^{3} 
    \frac{ \left[ \left(  M_{q}^{fit} \right)_{kl} - \left(  M_{q}^{th} \right)_{kl} \right]^{2} 
    }{ \left( \sigma_{ \left( M_{q} \right)_{kl} } \right)^{2}  },  
  \end{array}  
 \end{equation}  
 }
where the  matrix {$M_{q}^{fit}$} is given in eq.~(\ref{Md:valores}) or~(\ref{Mu:valores}),
 whereas the explicit form of {$M_{q}^{th}$} is obtained from eq.~(\ref{MMM}). 
 The matrix {$M_{q}^{th}$} has nine free parameters while {$M_{q}^{fit}$} has only four 
 independent parameters for use in the respective $\chi^{2}$ fit. So, we need to fix at least five parameters of
 {$M_{q}^{th}$} to obtain that the $\chi^{2}$ function does not contain degrees of freedom, however, in 
 this case we can only know the minimum value that the $\chi^{2}$ function takes. Thus, if we want to obtain 
 an allowed region with a certain confidence level, we must fix at least six parameters of 
 {$M_{q}^{th}$}.} In a previous discussion, we determined that the parameters 
 more sensitive to flavour violating effects are the off-diagonal terms of the Yukawa texture 
 {$\left( \widetilde{\chi}_{n}^{q} \right)_{ij}$} ($i \neq j$), which have to be constrained. 
 Therefore,
in order to be consistent with the flavour physics constraints  reported in ~\cite{HernandezSanchez:2012eg,HernandezSanchez:2013xj,Crivellin:2013wna}, we 
 eventually considered the  allowed regions for  the off-diagonal terms of the Yukawa texture given in eq. (29).
 Then, the only free parameters in the function  { $\chi^{2}_{ _{ M_q } }$} are 
 {$\left( \widetilde{\chi}_{2}^{q} \right)_{kk}$}. 
 We performed in fact several $\chi^{2}$ fits, with $\tan \beta = 2, 6, 15, 30$ and  90 GeV $ <m_{H^{\pm}} < 500 $ GeV, at $90\%$ Confidence Level (CL). All the results of the $\chi^2$-fit of Yukawa matrices   are consistent with the current experimental measurements of the elements of the $V_{{\rm CKM }}$ matrix. 
 The average values of  $\left( \widetilde{\chi}_{2}^{q}\right)_{kk}$ parameters ($k =1,2,3$) in the range $90 <m_{H^{\pm}} < 500 \, \textrm{GeV}$},  are shown in Tab.~\ref{cuad-1}. 
These allowed regions for the diagonal terms of the Yukawa matrices are complementary results to the studies of flavour physics constraints in the 2HDM-III with a four-zero Yukawa texture, which clarify the usefulness of the fit. Now we can isolate both diagonal and off-diagonal terms of the Yukawa matrices with four-zero texture, by considering the flavour physics constraints and the fit presented here.
\begin{figure}[htb]
\begin{center}
\includegraphics[scale=0.39 ]{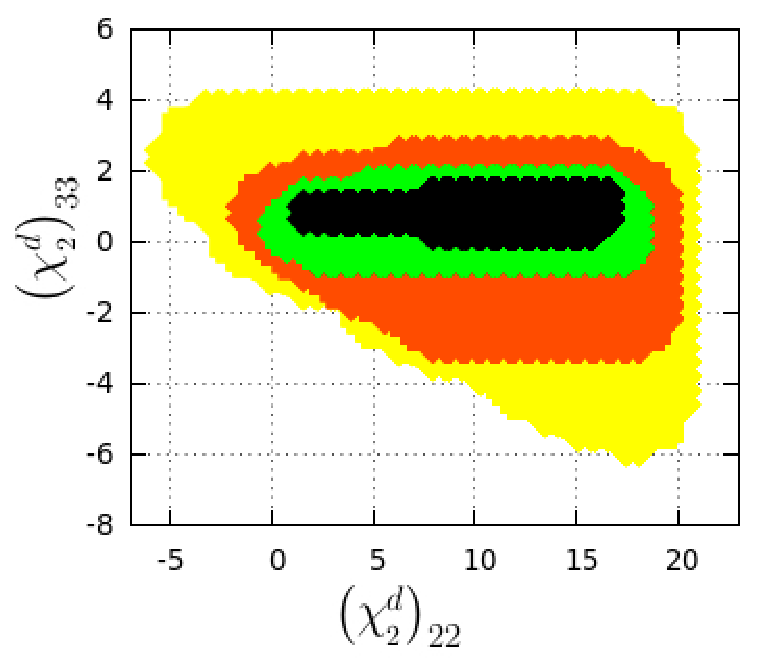} 
\includegraphics[scale=0.39 ]{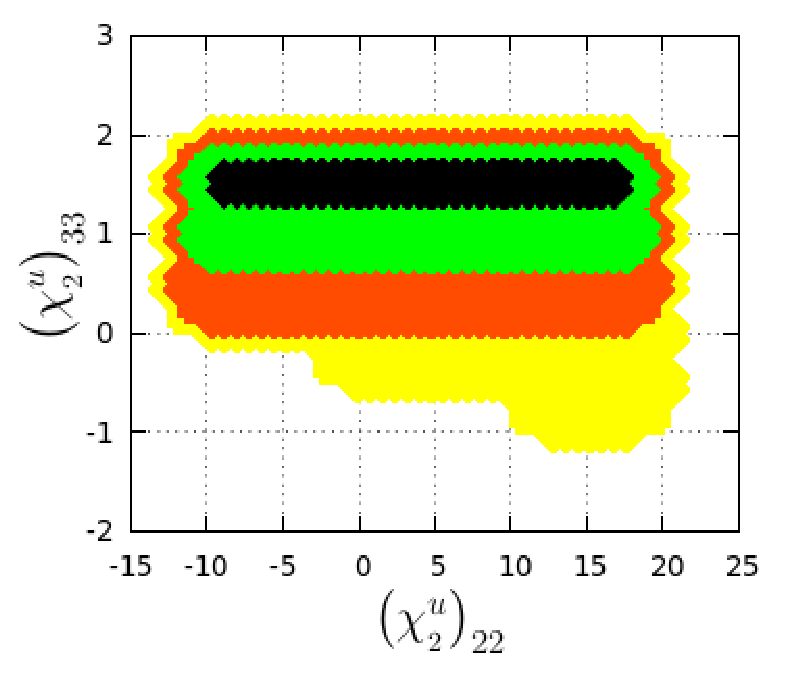}
\end{center}
\caption{ The allowed regions, at 90$\%$ CL, of the $\left( \widetilde{\chi}_{2}^{q}\right)_{kk}$ parameters ($k =2,3$) considering $\tan \beta = 2,\,6,\,15,\,30$, yellow, orange, green and black regions respectively.}
\protect{\label{Figura1}}
\end{figure}
The same permitted  regions for parameters space at $90\%$ CL are shown in Fig.~\ref{Figura1}. 
{\color{red}
{\scriptsize
%%%%%%%%%%Nuevas tablas
%
 \begin{table}
  \begin{center}
   \begin{tabular}{|c||c|c|c|c|} \hline \hline
  { $90 <\mu_{H^{+}} < 500 \, \textrm{GeV}$} &
  { $\tan \beta=2$ }& 
   { $\tan \beta=6$ }&  
   { $\tan \beta= 15$ }& 
   { $\tan \beta=30$ }\\ \hline 
    { $\left( \widetilde{\chi}_{2}^{u} \right)_{11}$ } & 
      {  $\left[ -3.92, 18 \right] $ } & 
      {  $\left[ 0.22 , 18 \right] $ } & 
      {  $\left[ 1.58 , 18 \right] $ }  & 
      {  $\left[ 3.76, 18 \right] $ } \\ \hline 
    { $\left( \widetilde{\chi}_{2}^{u} \right)_{22}$ } &
     { $\left[ -4.42 , 17.72 \right] $} & 
     { $\left[ -1.50 , 17.68 \right] $} &
     { $\left[ 0.50 , 17.46 \right] $} & 
     { $\left[ 2.84 , 16.24 \right] $} \\ \hline 
    { $\left( \widetilde{\chi}_{2}^{u} \right)_{33}$ } & 
     { $\left[ -0.84 , 1.98 \right] $}  & 
     { $\left[ -0.04 , 1.68 \right] $} &
     { $\left[  0.84 , 1.46 \right] $} &
     { $\left[  1.06 , 7.2 \right] $} \\ \hline 
    { $\left( \widetilde{\chi}_{2}^{d} \right)_{11}$ } & 
     { $\left[-2.74, 18 \right] $} & 
     { $\left[0.24, 18\right]$} &
     { $\left[3.6, 18\right]$} &
     { $\left[5.78, 18\right]$} \\ \hline 
    { $\left( \widetilde{\chi}_{2}^{d} \right)_{22}$ } & 
     { $\left[-2.66, 18.4 \right] $ } & 
     { $\left[0.82, 18.32 \right] $ } &
     { $\left[2.3, 17 \right] $} &   
    {  $\left[4.7, 15.72 \right] $} \\ \hline 
    { $\left( \widetilde{\chi}_{2}^{d} \right)_{33}$ } & 
     { $\left[-7.68, 4.22 \right] $ } & 
     { $\left[-2.28, 2.44 \right] $ } &
     { $\left[-0.45, 1.38\right] $ } &    
     { $\left[0.02, 1.0\right] $ } \\ \hline
    { $\chi^{2}_{u} \left( \chi^{2}_{d} \right)$ } & 
    { $1 \left( 0.21 \right)$ } & 
    { $1.17 \left( 0.30 \right)$ } & 
    { $1.26 \left( 0.50 \right)$ }  & 
    { $1.38 \left( 0.78 \right)$ }  \\ \hline \hline 
     \end{tabular}
   \caption{ Average values  of  $\left( \widetilde{\chi}_{2}^{q}\right)_{kk}$ parameters ($k =1,2,3$) in the range $90 <m_{H^{\pm}} < 500 \, \textrm{GeV}$, when the $\chi^2$-fit of Yukawa matrices is implemented.}    
  \label{cuad-1}
  \end{center} 
 \end{table} 
 }}
\section{Conclusions} \label{section concl}
 In conclusion, we  have presented the correlations 
 between the Yukawa matrices in the framework of the 2HDM, originating from a four-zero Yukawa texture, and 
 the current data on the CKM matrix through 
 a  $\chi^2$ fit. Firstly, we presented the diagonalisation of the Yukawa matrices, eventually showing that 
 the Cheng and Sher ansatz is a particular case of our general study. 
%Here, the complete analytical  
% expressions of the diagonalisation procedure  were obtained. 
Secondly, we have performed a numerical 
 analysis via  a $\chi^2$ fit  to the Yukawa matrices with respect to the measured entries of the CKM matrix. 
 As a consequence, we have obtained bounds for the parameters of the Yukawa texture, in particular for its 
 diagonal terms.  We have obtained results that are in complete 
 agreement with the  bounds obtained in our previous work in which we studied the flavour-violating 
 constraints. As an outlook, we deem our current parameterisation a more easily implementable one with 
 respect to our previous ones, thus we recommend its use for numerical analyses.
%%%%%%%%%%%%%%%%%%%%%%%%%%%%%%%%%%%%%%%%%%%%%%%%%%%
%%	A C K N O W L E D G M E N T S
%%%%%%%%%%%%%%%%%%%%%%%%%%%%%%%%%%%%%%%%%%%%%%%%%%%

\section*{Acknowledgments}
This work has been supported in part by\textit{ SNI-CONACYT (M\'exico)} and by \textit{PROMEP 
(M\'exico)} under the grant ``Red Tem\'atica: F\'isica del Higgs y del sabor''. FGC acknowledges the 
financial support from {\it CONACyT} and {\it PROMEP} under grants 208055 and 103.5/12/2548. SM is financed in part through the NExT Institute and is grateful to the University of Puebla for kind hospitality while parts of this work were being carried out.
%\end{acknowledgments}

\section*{References}

\bibliography{Referencias-fit}

\end{document}